\documentclass[10pt]{article}
\usepackage{latexsym}
\usepackage[dvips]{graphicx,color}
\newcommand{\be}{\begin{equation}}
\newcommand{\ee}{\end{equation}}
\def\n{\noindent}
\catcode `\@=11
\catcode `\@=12
\begin{document}
\begin{center}
\large{\bf {Universe With Time Dependent Deceleration Parameter and $\Lambda$ Term in 
General Relativity}} \\
\vspace{10mm}
\normalsize{Anirudh Pradhan\footnote{Corresponding author}} \\
\normalsize{\it{Department of Mathematics, Hindu Post-graduate College, Zamania-232 331, 
Ghazipur, India}} \\
\normalsize{\it{E-mail: pradhan@iucaa.ernet.in, acpradhan@yahoo.com}}\\
\vspace{10mm}
\normalsize{Saeed Otarod} \\
\normalsize{\it{Department of Physics, Yasouj University, Yasouj, Iran }} \\
\normalsize{\it{E-mail: sotarod@mail.yu.ac.ir, sotarod@yahoo.com}}\\
\end{center}
\vspace{10mm}
\begin{abstract} 
{A new class of exact solutions of Einstein's field equations with perfect 
fluid for an LRS Bianchi type-I spacetime is obtained by using a time dependent 
deceleration parameter. We have obtained a general solution of the field equations 
from which three models of the universe are derived: exponential, polynomial and 
sinusoidal form respectively. The behaviour of these models of the universe are 
also discussed in the frame of reference of recent supernovae Ia observations.}
\end{abstract}
\smallskip
\n PACS No. : 98.80.Es, 98.80.-k \\
\n Keywords : Cosmology, cosmological term, deceleration parameter.
\section{Introduction}
The Bianchi cosmologies play an important role in theoretical cosmology and 
have been much studied since the 1960s. A Bianchi cosmology represents a spatially 
homogeneous universe, since by definition the spacetime admits a three-parameter 
group of isometries whose orbits are spacelike hyper-surfaces. These models can be
used to analyze aspects of the physical Universe which pertain to or which may be 
affected by anisotropy in the rate of expansion, for example , the cosmic microwave 
background radiation, nucleosynthesis in the early universe, and the question of the 
isotropization of the universe itself \cite{ref1}. For simplification and description 
of the large scale behaviour of the actual universe, locally rotationally symmetric 
[henceforth referred as LRS] Bianchi I spacetime have widely studied 
{\cite{ref2}$-$\cite{ref6}}. When the Bianchi I spacetime expands equally in two 
spatial directions it is called locally rotationally symmetric. These kinds of models 
are interesting because Lidsey \cite{ref7} showed that they are equivalent to a flat 
(FRW) universe with a self-interesting scalar field and a free massless scalar field, 
but produced no explicit example. Some explicit solutions were pointed out in references 
\cite{ref8,ref9}.
\newline
\par
The Einstein's field equations are coupled system of high non-linear differential 
equations and we seek physical solutions to the field equations for their applications 
in cosmology and astrophysics. In order to solve the field equations we normally assume 
a form for the matter content or that spacetime admits killing vector symmetries \cite{ref10}. 
Solutions to the field equations may also be generated by applying a law of variation for 
Hubble's parameter which was proposed by Berman \cite{ref11}. In simple cases the Hubble 
law yields a constant value of deceleration parameter. It is worth observing that most of 
the well-known models of Einstein's theory and Brans-Deke theory with curvature parameter 
$k = 0$, including inflationary models, are models with constant deceleration parameter. 
In earlier literature cosmological models with a constant deceleration parameter have been 
studied by Berman \cite{ref11}, Berman and Gomide \cite{ref12}, Johri and Desikan \cite{ref13}, 
Singh and Desikan \cite{ref14}, Maharaj and Naidoo \cite{ref15}, Pradhan and {\it et al.} \cite{ref16} 
and others. But redshift magnitude test has had a chequered history. During the 1960s and the 1970s, 
it was used to draw very categorical conclusions. The deceleration parameter $q_{0}$ was then 
claimed to lie between $0$ and $1$ and thus it was claimed that the universe is decelerating. 
Today's situation, we feel, is hardly different. Observations \cite{ref17,ref18} of Type Ia 
Supernovae (SNe) allow to probe the expansion history of the universe. The main conclusion of 
these observations is that the expansion of the universe is accelerating. So we can consider 
the cosmological models with variable deceleration parameter. The readers are advised to see the 
papers by Vishwakarma and Narlikar \cite{ref19} and Virey {\it et al.} \cite{ref20} and references 
therein for a review on the determination of the deceleration parameter from Supernovae data.
\newline
\par
Motivated with the situation discussed above, in this paper we can focus upon the problem 
of establishing a formalism for studying the relativistic evolution for a time dependent 
deceleration parameter in an expanding universe. This paper is organized as follows. The 
metric and the field equations are presented in Section 2. In Section 3 we deal with a 
general solution. The Sections 4, 5, and 6 contain the three different cases for the solutions 
in exponential, polynomial and sinusoidal forms respectively. Finally in Section 7 concluding 
remarks will be given. 

\section{The Metric and Field  Equations}
We consider the LRS Bianchi type-I  metric in the form \cite{ref5}
\begin{equation}
\label{eq1}
ds^{2} = dt^{2} - A^{2} dx^{2} - B^{2}(dy^{2} + dz^{2}),
\end{equation}
where A and B are functions of $x$ and $t$.
The energy momentum-tensor in the presence of perfect fluid has the form 
\begin{equation}
\label{eq2}
T_{ij} = (\rho + p)u_{i}u_{j} - p g_{ij},  
\end{equation}
where $\rho$, $p$ are the energy density,
thermodynamical pressure respectively and $u_{i}$ is the four velocity  vector satisfying 
the relations
\begin{equation}
\label{eq3}
u_{i}u^{i} = 1,
\end{equation}
The Einstein's field equations (in gravitational units $c = 1$, $G = 1$) read as
\begin{equation}
\label{eq4}
R_{ij} - \frac{1}{2} R g_{ij} + \Lambda g_{ij} = -8\pi T_{ij},
\end{equation}
where $R_{ij}$ is the Ricci tensor; $R$ = $g^{ij} R_{ij}$ is the
Ricci scalar. The Einstein's field equations (\ref{eq4}) for the line element (\ref{eq1})
has been set up as
\begin{equation}
\label{eq5}
\frac{2\ddot{B}}{B} + \frac{\dot{B}^{2}}{B^{2}} - \frac{{B^{\prime}}^{2}}{A^{2}B^{2}} = 
- 8\pi p + \Lambda,
\end{equation}
\begin{equation}
\label{eq6}
{\dot{B}}^{\prime} - \frac{B^{\prime}\dot{A}}{A} = 0,
\end{equation}
\begin{equation}
\label{eq7}
\frac{\ddot{A}}{A} + \frac{\ddot{B}}{B} + \frac{\dot{A}\dot{B}}{AB} - \frac{B^{\prime \prime}}
{A^{2}B} + \frac{A^{\prime}B^{\prime}}{A^{3}B} = - 8\pi p + \Lambda,
\end{equation}
\begin{equation}
\label{eq8}
\frac{2 B^{\prime \prime}}{A^{2}B} - \frac{2 A^{\prime}B^{\prime}}{A^{3}B} + 
\frac{{B^{\prime}}^{2}}{A^{2}B^{2}} - \frac{2\dot{A}\dot{B}}{AB} - \frac{\dot{B}^{2}}{B^{2}}
= 8\pi \rho - \Lambda. 
\end{equation}
The energy conservation equation yields
\begin{equation}
\label{eq9}
\dot{\rho} + (p + \rho)\left(\frac{\dot{A}}{A} + \frac{2\dot{B}}{B}\right) + \dot{\Lambda} = 0,
\end{equation}
where dots and primes indicate partial differentiation with respect to $t$ and $x$ 
respectively. 

In order to completely determine the system, we choose a barotropic equation of state
\begin{equation}
\label{eq10}
p = \gamma \rho, 0 \leq \gamma \leq 1.
\end{equation}

\section{Solution of the Field Equations}
Equation (\ref{eq6}), after integration, yields
\begin{equation}
\label{eq11}
A = \frac{B^{\prime}}{\ell},
\end{equation}
where $\ell$ is an arbitrary function of $x$. Equations (\ref{eq5}) and (\ref{eq7}), 
with the use of Eq. (\ref{eq11}), reduces to 
\begin{equation}
\label{eq12}
\frac{B}{B^{\prime}}\frac{d}{dx}\left(\frac{\ddot{B}}{B}\right) + \frac{\dot{B}}{B^{\prime}}
\frac{d}{dt}\left(\frac{B^{\prime}}{B}\right) + \frac{\ell^{2}}{B^{2}}\left(1 - \frac{B}
{B^{\prime}}\frac{\ell^{\prime}}{\ell}\right) = 0.
\end{equation}
If we assume $\frac{B^{\prime}}{B}$ to be a function of $x$ alone, then $A$ and $B$ are 
separable in $x$ and $t$. Hence, after integrating Eq. (\ref{eq12}) gives
\begin{equation}
\label{eq13}
B = \ell S(t),
\end{equation}
where $S$ is a scale factor which is an arbitrary function of $t$. Thus from Eqs. (\ref{eq11}) 
and (\ref{eq13}), we have
\begin{equation}
\label{eq14}
A = \frac{\ell^{\prime}}{\ell} S.
\end{equation}
Now the metric (\ref{eq1}) is reduced to the form
\begin{equation}
\label{eq15}
ds^{2} = dt^{2} - S^{2}\left[dX^{2} + e^{2X}(dy^{2} + dz^{2})\right],
\end{equation}
where $X = ln~{\ell}$. The mass-density, pressure and Ricci scalar are obtained as
\begin{equation}
\label{eq16}
8\pi \rho = \frac{3}{S^{2}}\left[{\dot{S}}^2 - 1\right] + \Lambda,
\end{equation}
\begin{equation}
\label{eq17}
8\pi p = \frac{1}{S^{2}}\left[1 - {\dot{S}}^2 - 2\ddot{S}\right] - \Lambda,
\end{equation}
\begin{equation}
\label{eq18}
R = \frac{6}{S^{2}}\left[\ddot{S} + {\dot{S}}^2 - 1\right].
\end{equation}
The function $S(t)$ remains undetermined. To obtain its explicit dependence on $t$, 
one may have to introduce additional assumption. To achieve this, we assume the 
deceleration parameter to be variable, i.e.
\begin{equation}
\label{eq19}
q = - \frac{S\ddot{S}}{{\dot{S}}^2} = - \left(\frac{\dot{H} + H^{2}}{H^{2}}\right) = b (\rm variable),
\end{equation} 
where $H = \frac{\dot{S}}{S}$ is the Hubble parameter. The above equation may be rewritten as
\begin{equation}
\label{eq20}
\frac{\ddot{S}}{S}+ b \frac{{\dot{S}}^2}{S^{2}} = 0.
\end{equation} 
The general solution of Eq. (\ref{eq20}) is given by 
\begin{equation}
\label{eq21}
\int{e^{\int{\frac{b}{S}dS}}}dS = t + k,
\end{equation}
where $k$ is an integrating constant.

In order to solve the problem completely, we have to choose $\int{\frac{b}{S}dS}$ in such a 
manner so that Eq. (\ref{eq21}) be integrable. 

Let us consider
\begin{equation}
\label{eq22}
\int{\frac{b}{S}dS} = {\rm ln} ~ {L(S)},
\end{equation}
which does not effect the nature of generality of solution. Hence from Eqs. (\ref{eq21}) and 
(\ref{eq22}), one can obtain 
\begin{equation}
\label{eq23}
\int{L(S)dS} = t + k.
\end{equation}
Of course the choice of $L(S)$ is quite arbitrary but, since we are looking for physically 
viable models of the universe consistent with observations, we consider the following three 
cases:

\section{Solution in the Exponential Form}
Let us consider $ L(S) = \frac{1}{k_{1}S}$, where $k_{1}$ is an arbitrary constant.\\ 
In this case, on integrating, Eq. (\ref{eq23}) gives the exact solution
\begin{equation}
\label{eq24}
S(t) = k_{2}e^{k_{1}t},
\end{equation} 
where $k_{2}$ is an arbitrary constant. Using Eqs.(\ref{eq10}) and (\ref{eq24}) in 
Eqs. (\ref{eq16})-(\ref{eq18}), the mass-density, cosmological term and Ricci scalar are 
obtained as
\begin{equation}
\label{eq25}
\rho(t) = \left\{\frac{(3\gamma + 1)}{\gamma + 1} - \frac{3}{8\pi}\right\}\frac{1}{k_{2}^{2}
e^{2k_{1}t}} - \frac{2k_{1}^{2}}{(1 + \gamma)k_{2}}\frac{1}{e^{k_{1}t}} - 
\left\{\frac{(3\gamma + 1)}{\gamma + 1} - \frac{3}{8\pi}\right\}k_{1}^{2},
\end{equation} 
\begin{equation}
\label{eq26}
\Lambda(t) = \left\{\frac{1}{k_{2}^{2}e^{2k_{1}t}} - k_{1}^{2}\right\}\left(\frac{3\gamma}
{1 + \gamma}\right) - \frac{2k_{1}^{2}}{(1 + \gamma)k_{2}}\frac{1}{e^{k_{1}t}},
\end{equation}
\begin{equation}
\label{eq27}
R= \frac{6(k_{1}^{2}k_{2}e^{k_{1}t} - 1)}{k_{2}^{2}e^{2k_{1}t}} + 6k_{1}^{2}.
\end{equation}
\begin{figure}
\begin{center}
\includegraphics[angle=0, width=0.8\textwidth, height=0.35\textheight]{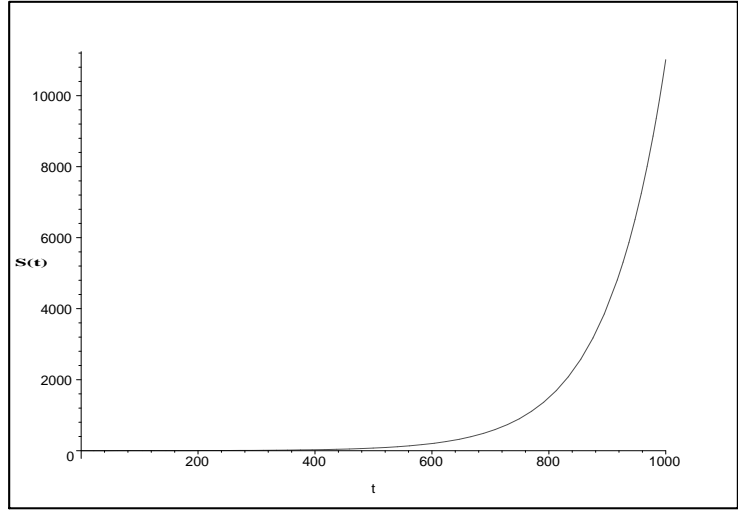}
\caption[DM delay tracks]{The plot of scale factor $S(t)$ vs time with parameters 
$k_1 = 0.01$, $k_2 = 0.5$, and $\gamma = 0.5$}
\label{fig:expscl}
\end{center}
\end{figure}

From Eq. (\ref{eq24}), since scale factor can not be negative, we find $S(t)$ is positive if 
$k_{2} > 0$. From Figure \ref{fig:expscl}, it can be deduced that at the early stages of the 
universe i.e. near $t = 0$, the scale factor of the universe had been approximately constant 
and had increased very slowly. At an specific time the universe has exploded suddenly and it 
has expanded to large scale. This fits nicely with Big Bang scenario.
\begin{figure}
\begin{center}
\includegraphics[angle=0, width=0.8\textwidth, height=0.35\textheight]{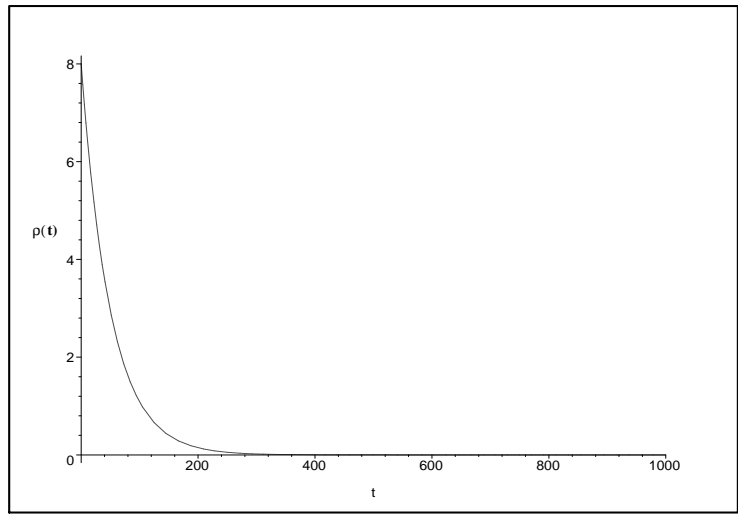}
\caption[DM delay tracks]{The plot of energy density $\rho(t)$ vs time with parameters 
$k_1 = 0.01$, $k_2 = 0.5$, and $\gamma = 0.5$}
\label{fig:exprho}
\end{center}
\end{figure}

From Eqs. (\ref{eq25}) and (\ref{eq26}), it is observed that $\rho(t) > 0$ and $\Lambda(t) > 0$ 
for $0 < t < \infty$ if $0 < k_{2} < 1$. Figure \ref{fig:exprho} clearly shows this 
behaviour of $\rho(t)$. 
\begin{figure}
\begin{center}
\includegraphics[angle=0, width=0.8\textwidth, height=0.35\textheight]{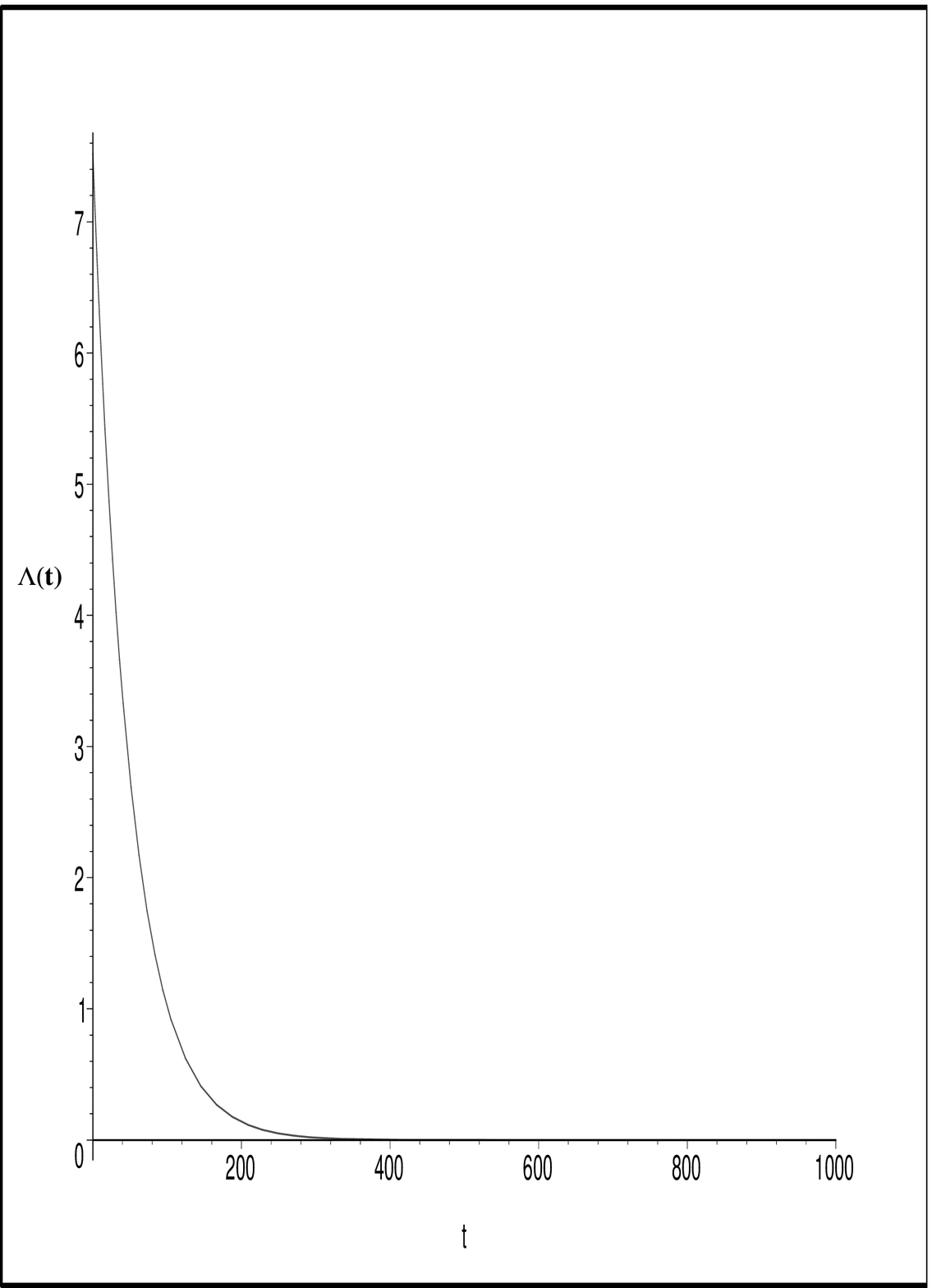}
\caption[DM delay tracks]{The plot of cosmological term ($\Lambda$) vs time with parameters 
$k_1 = 0.01$, $k_2 = 0.5$, and $\gamma = 0.5$}
\label{fig:explmd}
\end{center}
\end{figure}

From Eq. (\ref{eq26}), we observe that the cosmological term is a decreasing function of time and 
it approaches a small value as time progresses (i.e. the present epoch), which explains the small 
and positive value of $\Lambda$ at present (Perlmutter {\it et al.} \cite{ref21}; 
Riess {\it et al.} \cite{ref22}; Garnavich {\it et al.} \cite{ref23}; 
Schmidt {\it et al.} \cite{ref24}). Figure \ref{fig:explmd} clearly shows this 
behaviour of $\Lambda$ as decreasing function of time. From Eq. (\ref{eq27}), we see that 
the Ricci scalar remain positive for $$k_{1} > \frac{1}{\sqrt{k_{2}(1 + k_{2})}}.$$

\section{Solution in the Polynomial Form}
Let $ L(S) = \frac{1}{2k_{3}\sqrt{S + k_{4}}}$, where $k_{3}$ and $k_{4}$ are  constants. \\ 
In this case, on integrating, Eq. (\ref{eq23}) gives the exact solution
\begin{equation}
\label{eq28}
S(t) = \alpha_{1}t^{2} + \alpha_{2}t + \alpha_{3},
\end{equation} 
where $\alpha_{1}$, $\alpha_{2}$ and $\alpha_{3}$ are arbitrary constants. Using 
Eqs. (\ref{eq10}) and (\ref{eq28}) in Eqs. (\ref{eq16})-(\ref{eq18}), the mass-density, 
cosmological term and Ricci scalar are obtained as
\begin{equation}
\label{eq29}
\rho(t) = \frac{[(1 + 3\gamma) - (1 + 3\gamma)(2\alpha_{1}t + \alpha_{2})^{2} - 4\alpha_{1}]}
{(1 + \gamma)(\alpha_{1}t^{2} + \alpha_{2}t + \alpha_{3})^{2}},
\end{equation}
\begin{equation}
\label{eq30}
\Lambda(t) = \frac{[(2\alpha_{1}t + \alpha_{2})^{2} - 2\alpha_{1} - 1]}
{4\pi(1 + \gamma)(\alpha_{1}t^{2} + \alpha_{2}t + \alpha_{3})^{2}},
\end{equation}
\begin{equation}
\label{eq31}
R = \frac{6[(2\alpha_{1}t + \alpha_{2})^{2} + 2\alpha_{1} - 1]}
{(\alpha_{1}t^{2} + \alpha_{2}t + \alpha_{3})^{2}}.
\end{equation}
\begin{figure}
\begin{center}
\includegraphics[angle=0, width=0.8\textwidth, height=0.35\textheight]{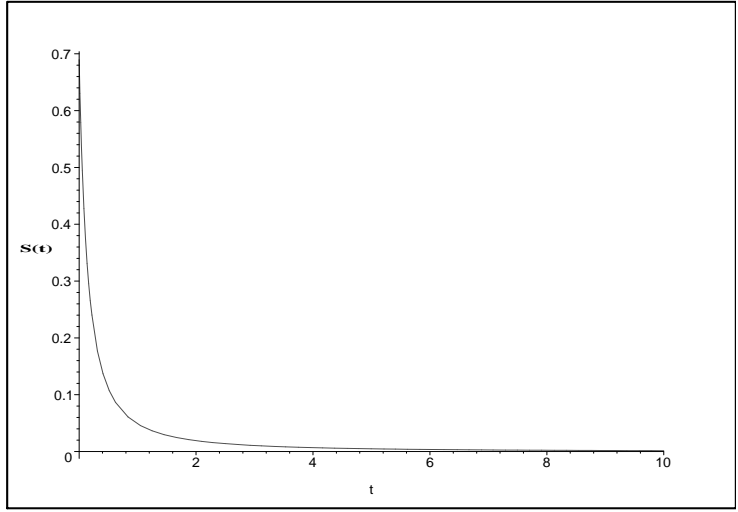}
\caption[DM delay tracks]{The plot of scale factor $S(t)$ vs time with parameters 
$\alpha_1 = 1.00$, $\alpha_2 = 4.00$, $\alpha_3 = 1.00$ and $\gamma = 0.5$}
\label{fig:polyscl}
\end{center}
\end{figure}

From Eq. (\ref{eq28}), it is observed that $S(t) > 0$ for $0 \leq t < \infty$ if 
$\alpha_{1}$, $\alpha_{2}$ and $\alpha_{3}$ are positive constants. From Figure $4$, 
it is observed that the scale factor is a decreasing function of time which means that 
our universe is expanding.    
\begin{figure}
\begin{center}
\includegraphics[angle=0, width=0.8\textwidth, height=0.35\textheight]{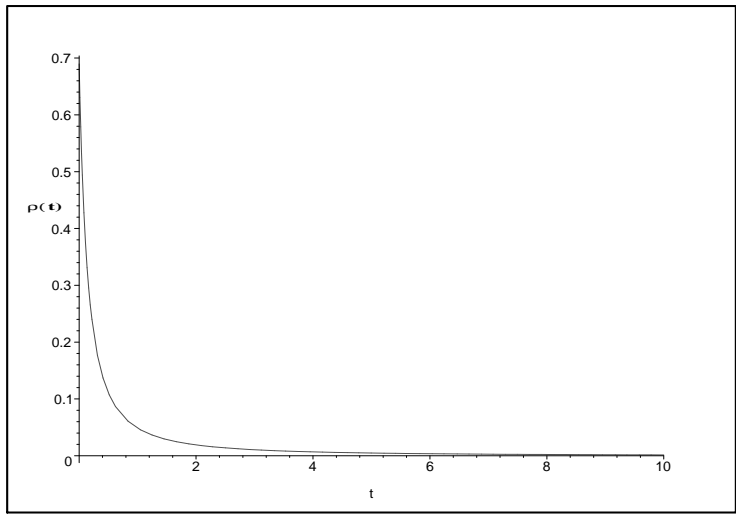}
\caption[DM delay tracks]{The plot of energy density $\rho(t)$ vs time with parameters 
$\alpha_1 = 1.00$, $\alpha_2 = 4.00$, $\alpha_3 = 1.00$ and $\gamma = 0.5$}
\label{fig:polyrho}
\end{center}
\end{figure}

In order to have $\rho > 0$ for $0 \leq t < \infty$, we must have $\alpha_{2}^{2} > 
2\alpha_{1} + 1$. From Eq. (\ref{eq31}) we observe that Ricci scalar remains positive 
if $\alpha_{2}^{2} > \frac{1 - 2\alpha_{1}}{6}$. This condition also implies that 
$\alpha_{1} < \frac{1}{2}$. Figure $5$ clearly shows the decreasing behaviour of $\rho(t)$ 
as time increases and is always positive. The interesting point is that all physical 
parameters in our models are defined at $t = 0$ and we do not have any singularity.  
\begin{figure}
\begin{center}
\includegraphics[angle=0, width=0.8\textwidth, height=0.35\textheight]{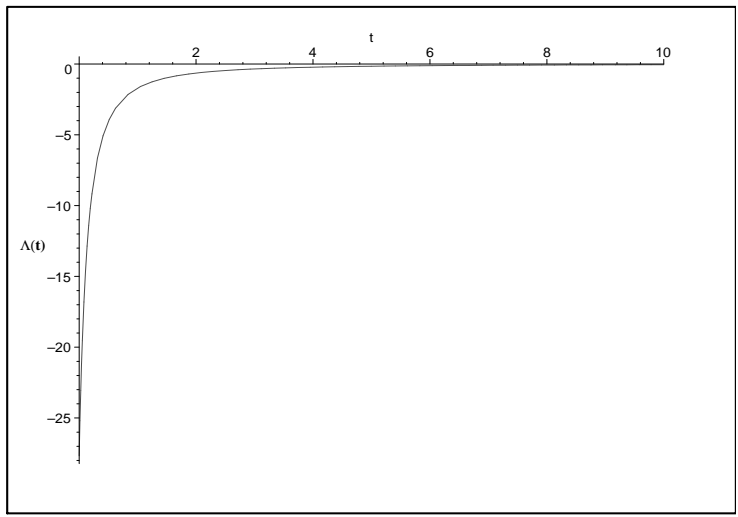}
\caption[DM delay tracks]{The plot of cosmological term ($\Lambda$) vs time with parameters 
$\alpha_1 = 1.00$, $\alpha_2 = 4.00$, $\alpha_3 = 1.00$ and $\gamma = 0.5$}
\label{fig:polylmd}
\end{center}
\end{figure}

It is observed from Eq. (\ref{eq30}) that $\Lambda(t)$ remains always negative but 
decreasing function of time. From the Figure \ref{fig:polylmd} it can be seen the 
behaviour of $\Lambda$ as a decreasing function of time. By decreasing we mean its 
absolute magnitude approaches zero which is acceptable physically. A negative 
cosmological term  adds to the attractive gravity of matter; therefore, universe 
with a negative cosmological term is invariably doomed to recollapse. A positive 
cosmological term resists the attractive gravity of matter due to its negative pressure. 
For most universe cosmological term eventually dominates over the attraction of matter 
and drives the universe to expands exponentially.

\section{Solution in the Sinusoidal Form}
If we set $ L(S) = \frac{1}{\beta\sqrt{1 - S^{2}}}$, where $\beta$ is constant. \\
In this case, on integrating, Eq. (\ref{eq23}) gives the exact solution
\begin{equation}
\label{eq32}
S = M\sin(\beta t) + N\cos(\beta t) + \beta_{1},
\end{equation} 
where $M$, $N$ and $\beta_{1}$ are constants. Using Eqs.(\ref{eq10}) and (\ref{eq32}) in 
Eqs. (\ref{eq16})-(\ref{eq18}), the mass-density, cosmological term and Ricci scalar are 
obtained as
\begin{equation}
\label{eq33}
4\pi(1 + \gamma)\rho = \frac{\left[\beta^{2}\left(M\cos(\beta t) - N\sin(\beta t)\right)^{2} + 
\beta^{2}\left(M\sin(\beta t) - N\cos(\beta_{1}t)\right) - 1 \right]}{\left(M\sin(\beta t)
+ N\cos(\beta t) + \beta_{1}\right)^{2}}
\end{equation}
\begin{equation}
\label{eq34}
(1 + \gamma)\Lambda = - \frac{\left[(1 + 3\gamma)\beta^{2}\left(M\cos(\beta t) - N\sin(\beta t)
\right)^{2} - (1 + 3\gamma) - 2k_{1}^{2}\left(M\sin(\beta t) + N\cos(\beta t)\right)\right]}
{\left(M\sin(\beta t) + N\cos(\beta t) + \beta_{1}\right)^{2}} 
\end{equation}
 \begin{equation}
\label{eq35}
R = \frac{6\left[\left(M\beta \cos(\beta t) - N\beta \sin(\beta t)\right)^{2} - \beta^{2}
\left(M\sin(\beta t) + N\cos(\beta t)\right) - 1\right]}{\left(M\sin(\beta t) + N\cos(\beta t) 
+ \beta_{1}\right)^{2}}
\end{equation}
Since, in this case, we have many alternatives for choosing values of $M$, $N$, $\beta$, 
$\beta_{1}$, it is just enough to look for suitable values of these parameters, such that 
the physical initial and boundary conditions are satisfied. We are trying to find feasible 
interpretation and situations relevant to this case. Further study in this case is in progress.
\begin{figure}
\begin{center}
\includegraphics[angle=0, width=0.8\textwidth, height=0.35\textheight]{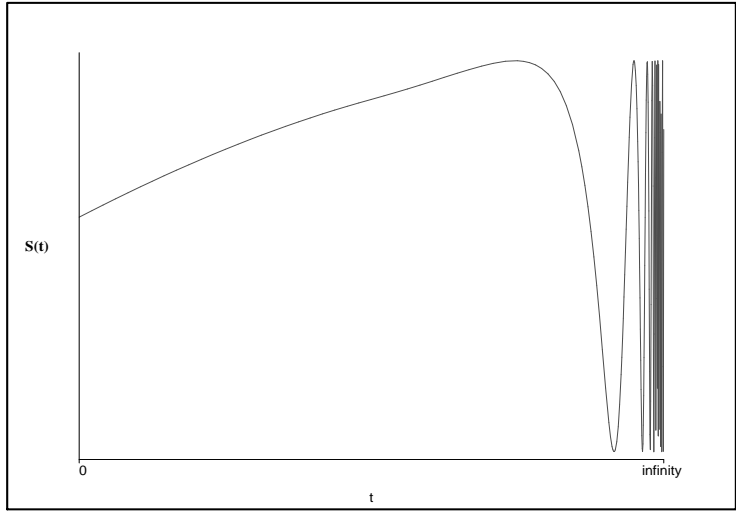}
\caption[DM delay tracks]{The plot of scale factor $S(t)$ vs time with parameters 
$M = 2.00$, $N = 1.00$, $\beta = 10.00$, $\beta_1 = 0.2$, and $\gamma = 0.5$}
\label{fig:sinscl}
\end{center}
\end{figure}

From the Figure \ref{fig:sinscl} it is observed that at early stages of the universe, 
the scale of the universe increases gently and then decreases sharply, and afterwords 
it will oscillate for ever.
\begin{figure}
\begin{center}
\includegraphics[angle=0, width=0.8\textwidth, height=0.35\textheight]{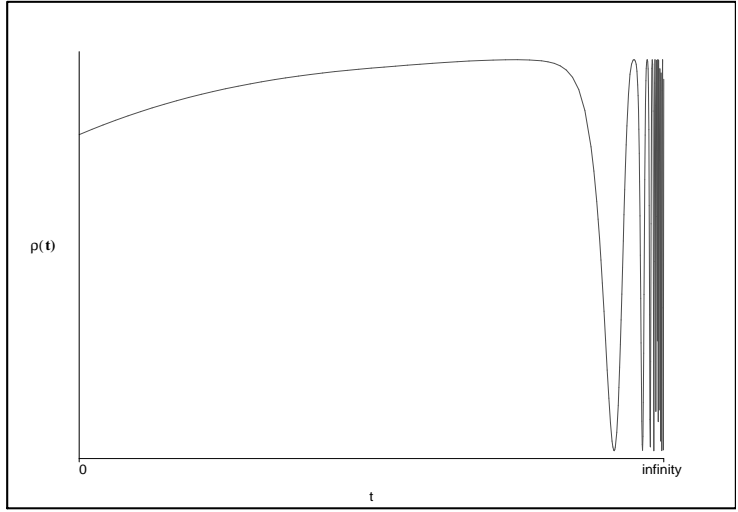}
\caption[DM delay tracks]{The plot of energy density $\rho(t)$ vs time with parameters 
$M = 2.00$, $N = 1.00$, $\beta = 10.00$, $\beta_1 = 0.2$, and $\gamma = 0.5$}
\label{fig:sinrho}
\end{center}
\end{figure}
\begin{figure}
\begin{center}
\includegraphics[angle=0, width=0.8\textwidth, height=0.35\textheight]{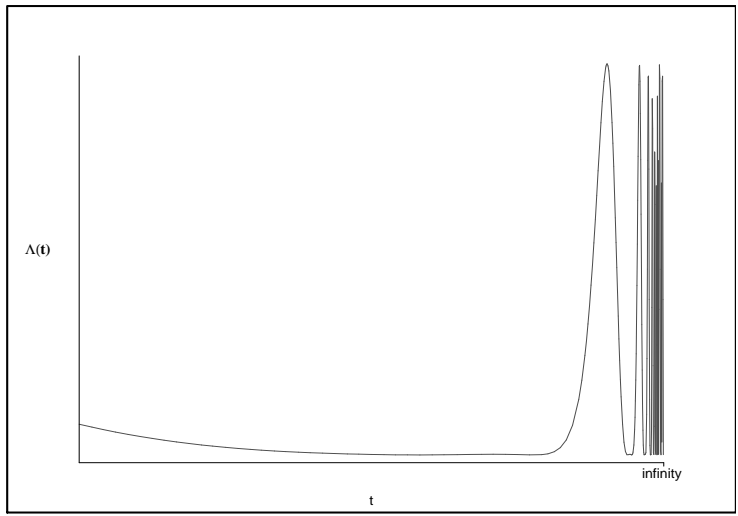}
\caption[DM delay tracks]{The plot of cosmological term ($\Lambda$) vs time with parameters 
$M = 2.00$, $N = 1.00$, $\beta = 10.00$, $\beta_1 = 0.2$, and $\gamma = 0.5$}
\label{fig:sinlmd}
\end{center}
\end{figure}

From Figure \ref{fig:sinrho} and Figure \ref{fig:sinlmd} we conclude that at early stages 
of the universe the matter is created as a result of loss of vacuum energy and at a 
particular epoch it has started to oscillate for ever due to sinusoidal property. It is 
worth to mention here that in this case oscillation takes place in positive quadrant which 
has physical meaning. 
\section{Conclusions}
In this paper we have described a new class of LRS Bianchi type I cosmological models 
with a perfect fluid as the source of matter by applying a variable deceleration 
parameter. Generally, the models are expanding, non-shearing and isotropic in nature.

The cosmological term in the model in Section 4 is a decreasing function of time and this 
approaches a small value as time increases (i.e. present epoch). The value of the 
cosmological ``term'' for this model is found to be small and positive which is supported 
by the results from recent supernovae observations obtained by the High-Z Supernova 
Team and Supernova Cosmological Project (Perlmutter {\it et al.} \cite{ref21}; Riess {\it et al.} 
\cite{ref22}; Garnavich {\it et al.} \cite{ref23}; Schmidt {\it et al.} \cite{ref24}). The 
cosmological term in the model in Section 5 is also a decreasing function of time but it is 
always negative. A negative cosmological term  adds to the attractive gravity of matter; 
therefore, universe with a negative cosmological term is invariably doomed to recollapse. 
The cosmological term in Section 6 also decreases while time increases to a specific instant.
During this period as we can understand from Figure \ref{fig:sinlmd}, we will have enough 
matter creation to force the universe to oscillate for ever due to sinusoidal property of $\Lambda$. 
This means we always have annihilation and creation of matter permanently. At this point one more 
sentence may be added to our discussion and i.e. as the graphs for $\Lambda$ and $\rho$ in this 
case the explosion of the universe at the early stages of its creation has been only a consequence 
of matter creation.    
\section*{Acknowledgements} 
The authors thank to the Inter-University Centre for Astronomy and Astrophysics, 
India for providing  facility where this work was carried out. S. Otarod also 
thanks to the Yasouj University for providing leave during this visit to IUCAA.\\
\newline
\newline


\noindent
\begin{thebibliography}{99}
\bibitem {ref1} 
M. A. H. MacCallum, in {\it General Relativity: An Einstein Centenary Survey}, edited by
S. W. Hawking and W. Israel (Cambridge University Press, Cambridge, 1979). 
\bibitem {ref2}
J. Hajj-Boutros and J. Sfeila, {\it Int. J. Theor. Phys.} {\bf 26}, 98 (1987).
\bibitem {ref3}
Shri Ram, {\it Int. J. Theor. Phys.} {\bf 28}, 917 (1989).
\bibitem {ref4}
A. Mazumder, {\it Gen. Rel. Grav.} {\bf 26}, 307 (1994).
\bibitem {ref5}
A. Pradhan, K. L. Tiwari and A. Beesham, {\it Indian J. Pure Appl. Math.} {\bf 32}, 789 (2001). \\
A. Pradhan and A. K. Vishwakarma, {\it SUJST} {\bf XII} Sec. B, 42 (2000). \\
I. Chakrabarty and A. Pradhan, {\it Grav. \& Cosm.} {\bf 7}, 55 (2001). \\
A. Pradhan and A. K. Vishwakarma, {\it Int. J. Mod. Phys.} D {\bf 8}, 1195 (2002). \\
A. Pradhan and A. K. Vishwakarma, {\it J. Geom. Phys.} {\bf 49}, 332 (2004).
\bibitem {ref6}
G. Mohanty, S. K. Sahu and P. K. Sahoo, {\it Astrophys. Space Sci.} {\bf 288}, 523 (2003). 
\bibitem {ref7}
J. E. Lidsey, {\it Class. Quant. Grav.} {\bf 9}, 1239 (1992). 
\bibitem {ref8}
J. M. Aguirregabiria, A. Feinstein and J. Ibanez, {\it Phys. Rev.} D {\bf 48}, 4662 (1993).
\bibitem {ref9}
J. M. Aguirregabiria and L. P. Chimento, {\it Class. Quant. Grav.} {\bf 13}, 3197 (1966).
\bibitem {ref10}
D. Kramer, H. Stephani, M. MacCallum and E. Hertt, {\it Exact Solutions of Einstein's Field Equations}, 
Cambridge University Press, Cambridge, 1980. 
\bibitem {ref11}
M. S. Berman, {\it Nuovo Cimento} B {\bf 74}, 182 (1983).
\bibitem {ref12}
M. S. Berman and F. M. Gomide, {\it Gen. Rel. Grav.} {\bf 20}, 191 (1988).
\bibitem {ref13}
V. B. Johri and K. Desikan, {\it Pramana - J. Phys.} {\bf 42}, 473 (1994).
\bibitem {ref14}
G. P. Singh and K. Desikan, {\it Pramana - J. Phys.} {\bf 49}, 205 (1997).
\bibitem {ref15}
S. D. Maharaj and R. Naidoo, {\it Astrophys. Space Sci.} {\bf 208}, 261 (1993). 
\bibitem {ref16}
A. Pradhan, V. K. Yadav and I. Chakrabarty, {\it Int. J. Mod. Phys.} D {\bf 10}, 339 (2001). \\
A. Pradhan and I. Aotemshi, {\it Int. J. Mod. Phys.} D {\bf 9}, 1419 (2002). 
\bibitem {ref17}
R. K. Knop {\it et al.}, {\it Astrophys. J.} {\bf 598}, 102 (2003).  
\bibitem {ref18}
A. G. Riess {\it et al.}, {\it Astrophys. J.} {\bf 607}, 665 (2004).  
\bibitem {ref19}
R. G. Vishwakarma and J. V. Narlikar, {\it Int. J. Mod. Phys.} D {\bf 14}, 345 (2005). 
\bibitem {ref20}
J. -M. Virey, P. Taxil, A. Tilquin, A. Ealet, C. Tao and D. Fouchez, {\it On the determination 
of the deceleration parameter from Supernovae data}, astro-ph/0502163 (2005).
\bibitem {ref21}
S. Perlmutter {\it et al.}, {\it Astrophys. J.} {\bf 483}, 565 (1997), (astro-ph/9608192); 
{\it Nature} {\bf 391}, 51 (1998), (astro-ph/9712212); {\it Astrophys. J.} {\bf 517}, 565 (1999), 
(astro-ph/9608192).
\bibitem {ref22}
A. G. Riess {\it et al.}, {\it Astron. J.} {\bf 116}, 1009 (1998); (astro-ph/9805201).
\bibitem {ref23}
P. M. Garnavich {\it et al.}, {\it Astrophys. J.} {\bf 493}, L53 (1998a), (astro-ph/9710123); 
{\it Astrophys. J.} {\bf 509}, 74 (1998b); (astro-ph/9806396).
\bibitem {ref24}
B. P. Schmidt {\it et al.}, {\it Astrophys. J.} {\bf 507}, 46 (1998), (astro-ph/9805200). 
\end{thebibliography}
\end{document}